\documentclass[%
aip,
amsmath,
amssymb,
reprint
]{revtex4-2}

\usepackage{graphicx}
\usepackage{dcolumn}
\usepackage{bm}
\usepackage[utf8]{inputenc}
\usepackage[T1]{fontenc}
\usepackage{mathptmx}
\usepackage{textcomp}
\usepackage{siunitx}
\usepackage{bigints}

\usepackage{hyperref}
\usepackage[switch]{lineno}
\usepackage[table,xcdraw,dvipsnames]{xcolor}
\usepackage{booktabs}
\usepackage{float}
\usepackage{color}
\usepackage{ulem}
\usepackage{xr} % or use \usepackage{}

\begin{document}

\title{Steerable current-driven emission of spin waves in magnetic vortex pairs}

\author{Sabri~Koraltan}
\affiliation{Faculty of Physics, University of Vienna, Kolingasse 14-16, A-1090, Vienna, Austria}
\affiliation{Vienna Doctoral School in Physics, University of Vienna, A-1090, Vienna, Austria}
\affiliation{Research Platform MMM Mathematics - Magnetism - Materials, University of Vienna, Vienna 1090, Austria}
\email{sabri.koraltan@univie.ac.at}
\author{Katrin~Schultheiss}
\affiliation{Helmholtz-Zentrum Dresden-Rossendorf, Institute of Ion Beam Physics and Materials Research, 01328 Dresden, Germany}
\author{Florian~Bruckner}%
\affiliation{Faculty of Physics, University of Vienna, Kolingasse 14-16, A-1090, Vienna, Austria}
\author{Markus~Weigand}
\affiliation{Institut für Nanospektroskopie,Helmholtz-Zentrum Berlin für Materialien und Energie GmbH,
12489 Berlin, Germany}
\author{Claas~Abert}%
\affiliation{Faculty of Physics, University of Vienna, Kolingasse 14-16, A-1090, Vienna, Austria}
\affiliation{Research Platform MMM Mathematics - Magnetism - Materials, University of Vienna, Vienna 1090, Austria}
\author{Dieter~Suess}
\affiliation{Faculty of Physics, University of Vienna, Kolingasse 14-16, A-1090, Vienna, Austria}
\affiliation{Research Platform MMM Mathematics - Magnetism - Materials, University of Vienna, Vienna 1090, Austria}
\author{Sebastian~Wintz}%
\affiliation{Institut für Nanospektroskopie,Helmholtz-Zentrum Berlin für Materialien und Energie GmbH,
12489 Berlin, Germany}
\affiliation{Max Planck Institute for Intelligent Systems, 70569 Stuttgart, Germany}
\email{sebastian.wintz@helmholtz-berlin.de}

\date{\today}
\begin{abstract}
The efficient excitation of spin waves is a key challenge in the realization of magnonic devices. We demonstrate the current-driven generation of spin waves in antiferromagnetically coupled magnetic vortices. We employ time-resolved scanning transmission X-ray microscopy (TR-STXM) to directly image the emission of spin waves upon the application of an alternating current flowing directly through the magnetic stack. Micromagnetic simulations allow us to identify the origin of the excitation to be the current-driven Oersted field, which in the present system proves to be orders of magnitude more efficient than the commonly used excitation via stripline antennas. Our numerical studies also reveal that the spin-transfer torque can lead to the emission of spin waves as well, yet only at much higher current amplitudes. By using magnetostrictive materials, we futhermore demonstrate that the direction of the magnon propagation can be steered by increasing the excitation amplitude, which modifies the underlying magnetization profile through an additional anisotropy in the magnetic layers. The demonstrated methods allow for the efficient and tunable excitation of spin waves, marking a significant advance in the generation and control of spin waves in magnonic devices.
\end{abstract}

\maketitle

%\tableofcontents

\section*{Introduction}
\label{sec:introduction}
State-of-the-art data processing is nowadays based on complementary metal oxide semiconductor (CMOS) technology, where the control of current flow in transistors is used for logic operations and routing of information \cite{doyle2003high,manipatruni2018beyond}.
Accompanied by memory units such as random access memory (RAM) for temporary purposes or flash memory for persistent data storage, they build the backbone of today's computer designs. However, the relatively high power consumption of CMOS devices, due to ohmic losses, volatile refresh memories, and limits for further miniaturization, represents severe challenges for developing a sustainable information and communication technology for the future \cite{jacob2017scaling}.

Magnonic devices have emerged as a candidate solution to the challenges that CMOS is facing today~\cite{VVKruglyak2010,chumak2015magnon}. Within such devices, one makes use of \textit{magnons} (the quanta of spin waves) \cite{Bloch1930} to transfer low-loss information. As a local deflection of the magnetic orientation, spin waves typically exhibit wavelengths from subnanometers to centimeters at frequencies ranging from the megahertz to the terahertz range~\cite{chumak2015magnon, serga2010yig}. Based on their nonlinear propagation and coupling mechanisms, they can be used to operate various kinds of devices such as reprogrammable magnonic crystals~\cite{krawczyk2014review}, magnonic directional couplers~\cite{wang2020magnonic}, boolean computing devices~\cite{chumak2014magnon}, magnonic nano-ring resonators~\cite{wang2020nonlinear}, magnon memory devices~\cite{baumgaertl2023reversal}, magnonic logic circuits ~\cite{khitun2010magnonic}, offset-free magnetic field sensor~\cite{gattringer2023offset}, spin-wave majority gates~\cite{fischer2017experimental} or even multi-purpose devices ~\cite{ustinov2010multifunctional, wang_inverse-design_2021, papp2021nanoscale} and many more~\cite{chumak_advances_2022, barman20212021}. For real-world implementation of most proposed magnonic applications, it will be crucial to utilize spin waves of nanoscale wavelengths. This is mainly for two reasons: (i) miniaturization, where the wavelength imposes a constraint on the device footprint, and (ii) high group velocities in the short-wavelength (exchange-dominated) regime for sufficient processing speeds and propagation lengths. 

Typically, the excitation of coherent spin waves is achieved by using lithographically patterned microwave antennas, for example metallic coplanar waveguides \cite{mori2021broadband}. However, this method has limited efficiency, in particular for nanoscale wavelengths. Here, the wavelengths are restricted on the lower end to the minimum patterning sizes involved and at the same time such antennas still need to allow for sufficient impedance matching of the electric circuitry.
One way to overcome this limitation, among others, is to couple a global alternating external magnetic field, to local internal demagnetization fields \cite{Schlmann1964,Au2012,Yu2013} or to the dynamics of confined spin textures such as vortex cores~\cite{wintz2016magnetic, Behncke2018,Dieterle2019,Chang2020,Mayr2021,OsunaRuiz2021}, skyrmions ~\cite{mruczkiewicz2017spin, chen2021chiral, srivastava2023resonant, titze2023laser}, or domain walls~\cite{Hermsdoerfer2009,Whitehead2017,Hollnder2018,VandeWiele2016, wagner2016magnetic, henry2019unidirectional, che2023nonreciprocal,sluka2019emission}. The latter coupling to spin textures was found to be particularly resourceful in synthetic ferrimagnets (SFi), where two ferromagnetic layers are coupled antiparallelly through a non-ferromagnetic interlayer %~\cite{lan2017antiferromagnetic, qiu2022tunable}. 
\cite{Grnberg1986,Grnberg1981,wintz2016magnetic,Albisetti2020}. On the other hand, it was previously reported that oscillatory dynamics of spin textures can also be excited using the effects from electric currents flowing through these textures themselves, as a result of Oersted fields, spin-transfer torques, or spin-orbit torques \cite{Bolte2008,Locatelli2011,Litzius2016}. This can be the case for both direct and alternating currents as well as for lateral and vertical flow directions. 

\begin{figure*}
	\centering
	\includegraphics[width=\textwidth]{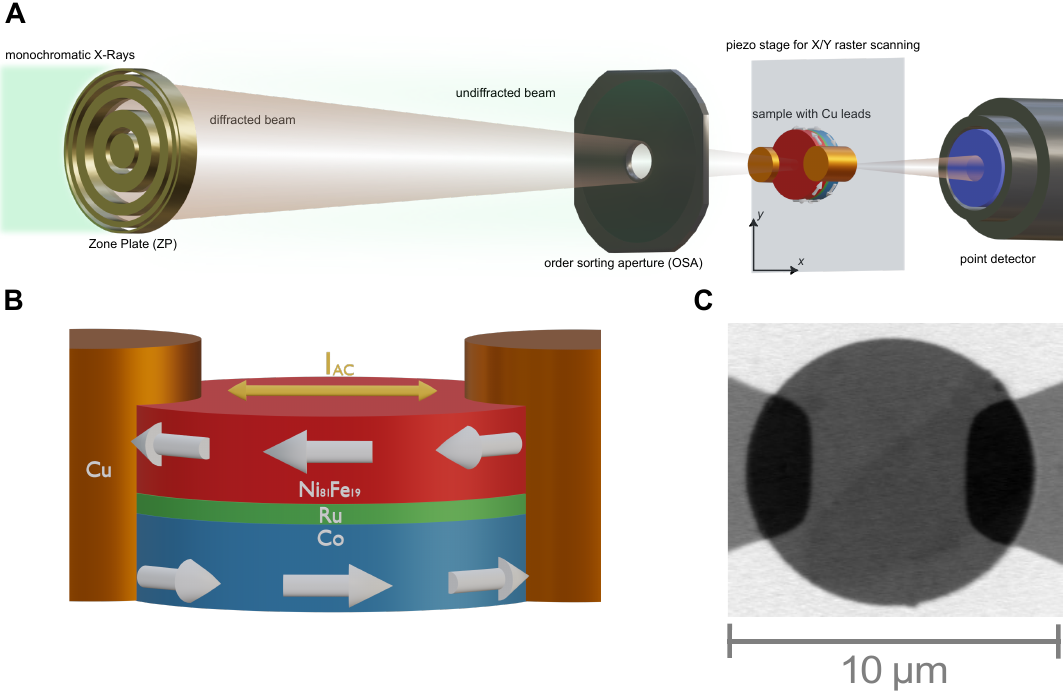}
 
	\caption{\textbf{Schematics of experimental setup and SFi Sample \#1.} (A) Scanning transmission X-ray microscopy setup, where the monochromatic X-rays are diffracted by the zone plate to focus in a single spot. The combination of the zone plate center stop and the order sorting aperture blocks the undiffracted beam and higher diffraction orders (latter not shown) from reaching the sample. The sample itself is grown on a thin membrane and mounted on a piezo stage that is raster scanned (x/y) through the focal point. The transmitted X-rays are collected by a point detector. (B) Schematic of the $\mathrm{Co/Ru/Ni_{81}Fe_{19}}$ microdisk with adjacent, partially overlapping copper leads (aspect ratios not to scale). White arrows indicate the general antiparallel orientation of the two ferromagnetic layers of the SFi. Alternating current injected from the leads flow laterally through the disk as indicated by the orange arrow. (C) Static STXM image of the sample recorded at the Fe $\mathrm{L_3}$ edge. The SFi microdisk (9 $\mu$m diameter) and the copper leads appear as gray contrast areas, with darker contrast in the overlapping regions. Within the disk, there is already faint magnetic contrast visible as well.}
	\label{fig:fig01}
\end{figure*}

Thus, it has been a key question, both in fundamental and practical respects, whether electric currents flowing through spin textures directly can be employed to generate short-wavelength spin waves and how efficient and versatile such a process would be as compared to the global field approach. Here, we demonstrate the realization of such current-driven spin-wave emission in SFi vortex pairs in a lateral alternating-current geometry and without the need for any magnetic bias field. We use high-resolution time-resolved X-ray microscopy to directly image the resulting spin-wave dynamics. Large-scale micromagnetic simulations allow us to understand the origin of the spin-wave emission, distinguish between different excitation mechanisms, and compare their efficiency. When the SFi consists of magnetostrictive materials, we demonstrate that the emission direction of the generated spin waves can be steered by the magnitude of the applied current.

\section*{Results}
\textbf{Current-driven spin-wave generation.} For our study, we require a time-resolved magnetic probe beyond the spatial resolution limit of visible-light techniques. Therefore, we image our samples using scanning transmission X-ray microscopy (STXM) with approximately 25 nm lateral resolution and by exploiting the X-ray magnetic circular dichroism \cite{schutz1987absorption} (XMCD) effect as magnetic contrast mechanism (see Methods). The typical STXM measurement setup is illustrated schematically in Fig.~\ref{fig:fig01}A. Measurements were carried out at the Maxymus end station \cite{weigand2022timemaxyne} at the BESSYII electron storage ring operated by the Helmholtz-Zentrum Berlin für Materialien und Energie. 
The first sample we examine is a circular thin film disk of 9 $\mu$m diameter made out of a Co($\SI{47.8}{nm}$)/Ru($\SI{0.8}{nm}$)/$\mathrm{Ni_{81}Fe_{19}}$($\SI{42.8}{nm}$) SFi stack (Sample \#1), as shown schematically in Fig.~\ref{fig:fig01}B. Here, we also highlight the electric contacting to the sample which is provided via copper thin film leads, overlapping at about 2 $\mu$m on either side with the microdisk. 
 A static STXM image of Sample \#1 recorded at the Fe $\mathrm{L_3}$ absorption edge is given in Fig.~\ref{fig:fig01}C. While the substrate background appears as bright contrast (high photon transmission), both the SFi microdisk and the copper leads appear as gray (medium transmission) or even dark gray where they overlap (low transmission). There is also already a hint for additional magnetic contrast within the disk, which we will turn to in the following.

\begin{figure*}
	\centering
	\includegraphics[width=\textwidth]{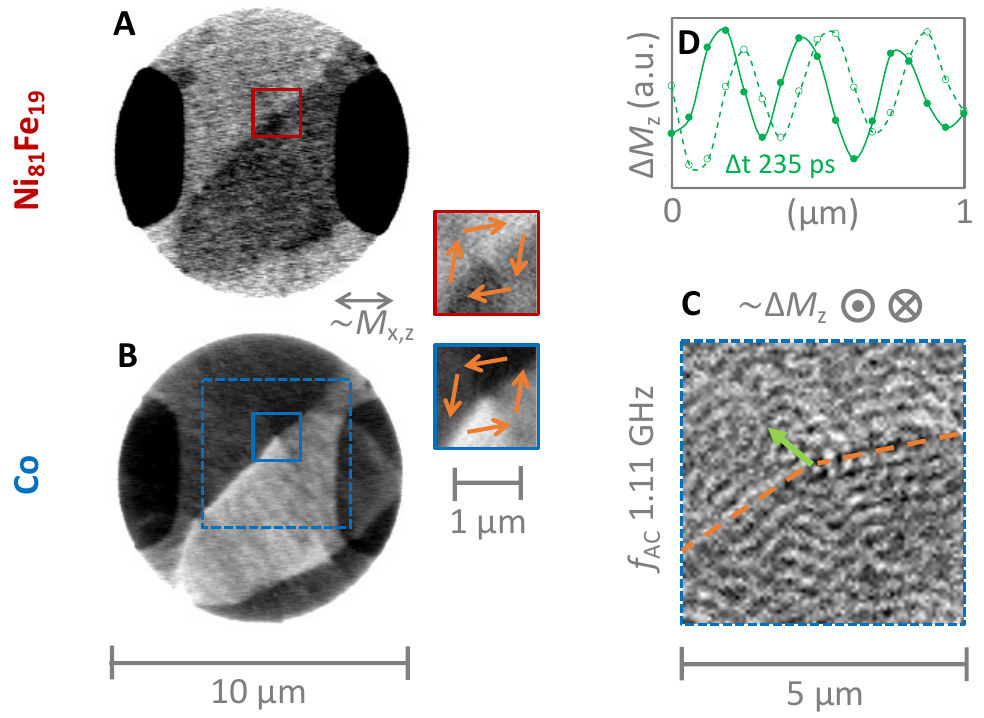}
	\caption{\textbf{STXM imaging of current-driven spin-wave emission in SFi Sample \#1.} (A,B) Static STXM images with partial in-plane magnetic sensitivity ($\sim M_x$), recorded at the Fe $\mathrm{L_3}$ edge (A) and the Co $\mathrm{L_3}$ edge (B), respectively. Zoom-in panels on the right correspond to the solid squares in (A) and (B), with orange arrows indicating the general magnetic vortex pair orientations as given by the black and white contrast. (C) Normalized TR-STXM snapshot of spin-wave dynamics in response to an alternating current of $f_{AC}=\SI{1.1}{GHz}$ flowing laterally through the disk, recorded with perpendicular magnetic sensitivity ($\sim \Delta M_z$) at the Co $\mathrm{L_3}$ edge. (D) Spin-wave amplitude profile extracted along the green arrow in (C) for a relative time delay of $\Delta t = 235$ ps.} 
	\label{fig:fig02}
\end{figure*}

To identify the magnetic state of Sample \#1, we performed static STXM imaging with partial in-plane magnetic sensitivity, separately for the two ferromagnetic layers. In addition to topographic information, Figures ~\ref{fig:fig02}A and B reveal the horizontal projection ($M_x$) of the magnetic orientation for the $\mathrm{Ni_{81}Fe_{19}}$ layer (recorded at Fe $\mathrm{L_3}$) and the Co layer (recorded at Co $\mathrm{L_3}$), respectively. By means of these images, it becomes obvious that the in-plane magnetic components of the two layers indeed are coupled antiparallelly (opposite relative contrast) and that there is a multivortex-domain state within the microdisk (multiple vortex cores in each layer). Despite this multidomain state, a large portion of the disk area can be identified as being part of a single interlayer vortex pair with antiparallel vortex circulations that have a slightly discrete and distorted rotation character (wall formation with rotation angles above and below $90^{\circ}$). The corresponding vortex cores are being somewhat displaced congruently to the upper right from the center of the disk. Zoom-in panels of the center regions (solid red and blue frames) allow for a closer view of the core area, with orange arrows illustrating the general vortex pattern. Although the core polarity cannot be directly obtained from these particular two images, it was found that typically the cores are aligned parallel in this kind of system \cite{wintz2016magnetic} (cf. Supplemental Fig.~\ref{fig:figs01}). Note that, in general, the STXM images of the Co layer have a higher signal-to-noise ratio than those of the $\mathrm{Ni_{81}Fe_{19}}$ layer, as there are many more absorbing Co atoms than Fe atoms in the sample, at a similar level of XMCD.

To evaluate whether current-driven spin-wave emission is possible in this kind of system, we send an alternating current laterally through Sample \#1, which is injected via the adjacent copper leads. The microwave current had a frequency of $f_{AC} = \SI{1.11}{GHz}$, a relevant current density amplitude of $j = 2.9 \times \SI{e10}{\mathrm{Am^{-2}}} $ and is flowing mainly along the $x$-axis, connecting the leads. Note that for the calculation of the current density we assume a uniform current flow in the SFi stack, where the cross-section area is defined as $A = l \times d$, where $l$ is the diameter of the disks ($l = \SI{9}{\mu m}$) and $d$ is the total thickness. Using TR-STXM (see Methods), we directly observe the dynamic response of the system. Figure ~\ref{fig:fig02}C shows a TR-STXM normalized snapshot in time $\Delta M_z(t_0)$, being sensitive to dynamic changes of the perpendicular component $M_z$ of the magnetization of the Co layer. The area represented by the dashed blue box in Fig.~\ref{fig:fig02}B is the corresponding image area. Indeed, we observe a clear spin-wave pattern of significant amplitude with an average wavelength of $307\pm30$ nm. By looking at consecutive snapshots in an animated way (Supplemental Movie M1), it becomes clear that these waves are mainly emitted from the somewhat bend, higher-angle domain walls (highlighted by orange dashed lines in Fig.~\ref{fig:fig02}C) and the vortex core. Note that the background magnetic state in Fig.~\ref{fig:fig02}C is 
 different from that in Fig.~\ref{fig:fig02} A/B with respect to the state of the domain wall (sample was remounted in between). The emission pattern can be seen in further detail in Supplemental Fig.~\ref{fig:figs01}, which displays the dynamic snapshots for the two layers separately in both absolute and normalized visualization, also revealing an in-phase dynamics of the $M_z$ component of the two layers. Thereby we can conclude, in line with earlier reports \cite{Grnberg1981,wintz2016magnetic}, that the observed spin waves are an instance of the acoustic, layer-collective SFi mode in the Damon-Eshbach geometry. Figure~\ref{fig:fig02}D further highlights the propagating nature of the observed spin waves by plotting the amplitude profile along the green arrow in Fig.~\ref{fig:fig02}C for a relative time delay of 235 ps and with additional lines for guiding the eye. The phase propagation is clearly visible. Note that the observed excitation process is not restricted to a certain frequency or resonance, but rather covers a broadband frequency range from hundreds of MHz to multiple GHz \cite{sluka2019emission} as shown exemplary in Supplemental Fig.~\ref{fig:figs02} and Supplemental Movie M2. Our experiments evidently demonstrate that short-wavelength spin-wave emission in spin textures can be driven by alternating currents flowing through these textures.
\newline

\textbf{Origin of the current-driven emission.}
In the following, we investigate the origin of the spin-wave emission from spin textures by employing micromagnetic simulations with \texttt{magnum.np}\cite{bruckner2023magnumnp}, see Methods for more details. 
When a charge current flows directly through a ferromagnetic material, several effects may occur to back-act on its magnetic state, two out of which are most prominent for our case. On the one hand, the charge current flowing along a given direction will generate an Oersted field within the conductor that is often axially symmetric. On the other hand, electrons that have acquired a spin polarization may be exerting a torque on the magnetization, termed the spin-transfer torque (STT) \cite{Berger1996,Slonczewski1996}. Both effects can be described by means of a micromagnetic continuum model \cite{abert2019micromagnetics}. Although, in reality, they are occurring at the same time, we can use micromagnetic simulations to apply them together, or individually, to investigate their role on the emission of spin waves. Thus, we perform such micromagnetic simulations, numerically solving the Landau-Lifshitz-Gilbert equation (LLG) in order to understand the origin of the spin-wave emission that we observed experimentally. Note that we chose a system analogous to Sample \#1, but limited the lateral size of the simulated disk to $4~\mu$m for computational practicability reasons.  Nevertheless, we confirmed that very similar results arise in even smaller disks, which validates our approach. Noteworthy, it is crucial to use double precision floating point operations to perform the numerical calculations in order to achieve the necessary accuracy.

\begin{figure*}
	\centering
	\includegraphics[width=\textwidth]{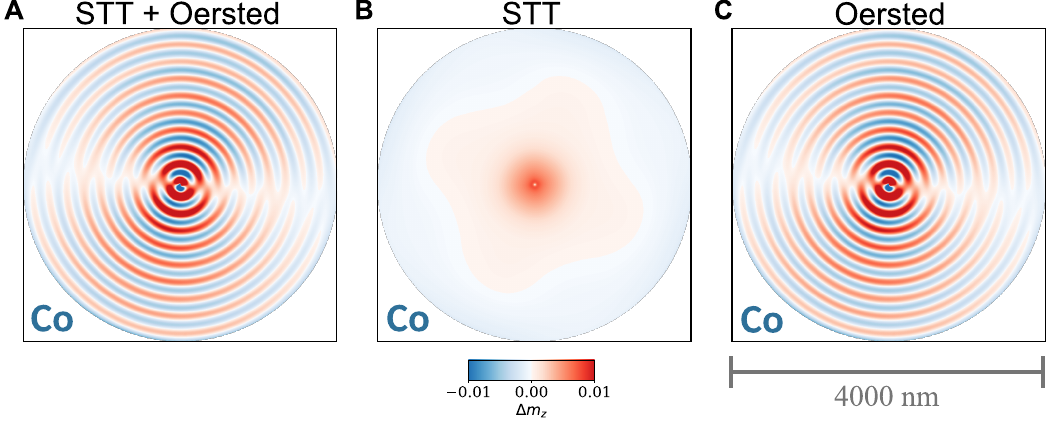}
	\caption{\textbf{Origin of current-driven spin-wave emission in SFi vortex pairs (simulations).} Snapshots of the normalized perpendicular dynamic magnetization ($\Delta m_{z}$) in the Co layer (averaged over the layer thickness) in a system analogous to Sample \#1 obtained from micromagnetic simulations with the contrast according to the blue-white-red color bar. The current density was modeled to $j = \SI{1e9}{\mathrm{Am^{-2}}}$ at $f_{AC} = \SI{1.11}{GHz}$. In (A), both the STT and the current-driven Oersted field were included in the effective field, while in (B) only the STT and in (C) only the Oersted field were considered, respectively.} 
	\label{fig:fig03}
\end{figure*}

Figure~\ref{fig:fig03} summarizes the different excitation methods simulated for an ideal vortex pair state with opposite circulations and parallel cores. It allows us to identify, in general, the main contributor to the current-driven emission of spin waves in our case. Figure~\ref{fig:fig03}A depicts the steady-state magnetization profile in the Co layer during the application of an AC current density of $j = \SI{1e9}{\mathrm{Am^{-2}}}$ and $f_{AC} = \SI{1.11}{GHz}$, where contributions from both the STT and the Oersted field are included in the modeling. For the sake of simplicity, we assume a uniform current flow along the $x$ axis throughout the stack. In reality, position dependent areal cross-sections, different resistivities of the materials and misalignments of the copper leads might induce inhomogeneities. One can see the emission of coherent, high-amplitude spin waves with wavelengths similar to those in the experiment. In contrast, there is no emission of spin waves anymore if one takes into consideration only the STT as shown in Fig.~\ref{fig:fig03}B. Even though the STT acts locally wherever the magnetization profile has a spatial gradient, the amplitude of the applied current is not sufficient to excite the magnetic vortex cores to gyrate in order to emit spin waves. If one considers only the generated Oersted field, then the resulting spin-wave pattern can be again observed in Fig.~\ref{fig:fig03}C, which is almost identical to the one in Fig.~\ref{fig:fig03}A. Therefore, we can conclude that it is solely the Oersted field that causes the efficient excitation of spin waves in our experiment. The magnetization dynamics only driven by Oersted fields is shown in the Supplemental Movie M3, and the Supplemental Movie M4 contains animations of spin-wave emissions for different excitation methods.
In addition, we provide the results from simulations with excitation current densities of the same order as in the experiment in Supplementary Fig.~\ref{fig:figs03}, where the magnetization and the normalized perpendicular dynamic magnetization are displayed separately for the Co and the $\mathrm{Ni_{81}Fe_{19}}$ layer, respectively.  These simulations demonstrate that the spin-wave amplitude increases with the applied current density, while we also observe a noticeable double-frequency ($2f$) excitation of spin waves at the same time.

\textbf{Excitation efficiency.} As we identified the Oersted field as the origin of the current-driven spin-wave emission, we further evaluate the excitation efficiency of this method. To this end, we compare the current-driven Oersted field-dominated spin-wave emission, to the previously used stripline antenna driven spin-wave emission from vortex cores\cite{wintz2016magnetic, sluka2019emission}, and also to the STT excitation above, yet considering two orders of magnitude higher current densities.

\begin{figure*}
	\centering
	\includegraphics[width=\textwidth]{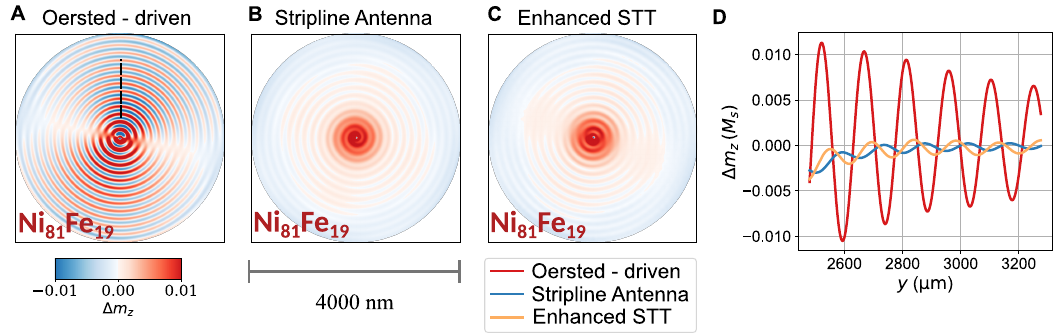}
	\caption{\textbf{Comparison of the spin-wave excitation efficiency in SFi vortex pairs (simulations).} Averaged spin-wave amplitude maps of the $\mathrm{Ni_{81}Fe_{19}}$ layer, given in terms of $\Delta m_z$ snapshots during excitation with (A) current-driven Oersted field, (B) uniform field of equal magnitude $max(H^{\mathrm{Oersted}}_y) = H^{\mathrm{uniform}}_y = \SI{400}{Am^{-1}}$ (as if the magnetic field was generated by a stripline antenna), and (C) enhanced STT excitation with a 100-fold increased current density.  The excitation frequency is $f = \SI{2}{GHz}$. The spin-wave amplitude along the black dotted line depicted in (A) is plotted in (D) for the three different cases according to the color code below (A).} 
	\label{fig:fig04}
\end{figure*}

Figure ~\ref{fig:fig04} gives an overview of the spin-wave amplitude profiles obtained from our micromagnetic simulations; this time at a frequency of $f_\mathrm{AC} = 2$ GHz. The current amplitude is again $j_{AC} = \SI{1e9}{\mathrm{A/m^{2}}}$. In contrast to Fig.~\ref{fig:fig03}, we now depict the magnetization of the $\mathrm{Ni_{81}Fe_{19}}$ layer, yet, as shown above, the Co layer behaves equivalently in terms of the perpendicular spin-wave component. The color-coded change in the normalized dynamic $\Delta m_{z}-$component of the magnetization, this time averaged over the $\mathrm{Ni_{81}Fe_{19}}$ layer, is strongest if the spin waves are excited via the current-driven Oersted field method, as can be seen in Fig.~\ref{fig:fig04}A. As referred to above, two other cases of spin-wave excitation were modeled numerically. The first involves a stripline antenna situated underneath the vortex pair, which produces an Oersted field that acts on the SFi (Fig.~\ref{fig:fig04}B). The second case involves only the STT from the current flowing through the sample, yet increasing the injected current density by a factor of $100$ (Fig.~\ref{fig:fig04}C). Both of these methods can excite spin waves, but the current-driven Oersted field generates spin waves with much higher amplitudes, making it the most effective excitation method of the three cases discussed, see also Supplemental Movie M4. The excitation efficiency can be further highlighted by means of the amplitude profiles in Fig.~\ref{fig:fig04}D. Note that the stripline antenna is assumed to generate a uniform field of equal magnitude compared to the current-driven Oersted field $ H^{\mathrm{uniform}}_y =\mathrm{max}(H^{\mathrm{Oersted}}_y) = \SI{20.9}{Am^{-1}}$. Comparing the spin-wave excitation efficiency of the three methods, we see that the amplitude of the spin waves that were excited by current-driven excitation is a factor of $\sim 10$ higher compared to the enhanced STT excitation, and a factor of $\sim 30$ higher compared to the excitation with the stripline antenna. In terms of energy coupling efficiency (considering the square of the amplitude), it means that current-driven Oersted field excitation has an impressive benefit of three orders of magnitude for the case of the given system. \newline

\begin{figure}
	\centering
	\includegraphics[width=\columnwidth]{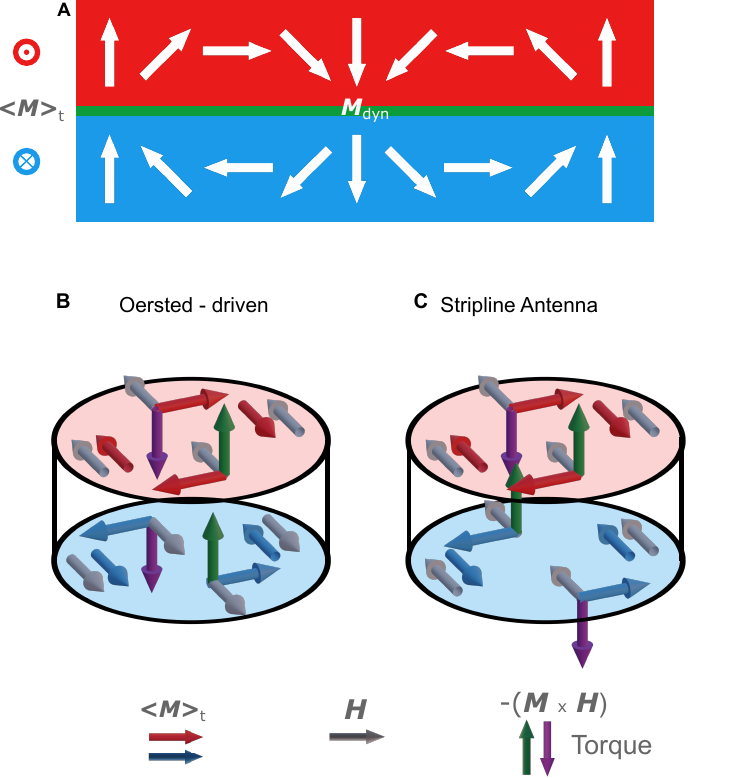}
	\caption{\textbf{SFi spin-wave profile and excitation symmetries (schematics).} (A) Cross-sectional profile of the dynamic magnetization components (white arrows) of an acoustic Damon-Eshbach spin wave in the SFi system with colors (blue and red). Note that here acoustic means that the perpendicular dynamic components are in-phase (in contrast to the dynamic in-plane components which are anti-phased). (B,C) Equilibrium magnetization in the antiparallel SFi vortex pairs (blue and red arrows) in the respective layers (blue and red disks), Oersted fields (gray arrows) at a snapshot in time for the Oersted-driven case (B) and stripline-antenna case (C). The locally resulting torque $(-\mathbf{M} \times \mathbf{H})$ is depicted by green (up) and purple (down) arrows. From symmetry perspective, only the Oersted driven excitation can directly excite the acoustic Damon-Eshbach SFi mode, as here the $z$-torques are in-phase between the two layers (B), in contrast to the stripline antenna excitation, where they are antiphased (C).}
 \label{fig:fig05}
\end{figure}

\textbf{Mechanisms of excitation.}The reason why the spin-wave generation efficiency is much higher for the current-driven case than for the uniform (stripline) case presumably lies in the symmetry of the excited spin-wave mode. This mode is the acoustic collective Damon-Eshbach mode of the SFi, which has in-phase perpendicular dynamic components and antiphase in-plane dynamic components for the two  ferromagnetic magnetic layers \cite{wintz2016magnetic}, Fig.~\ref{fig:fig05}A. Since the two layers have antiparallel in-plane equilibrium magnetic orientations, this mode cannot be directly excited by means of uniform in-plane fields as a consequence of symmetry considerations, see Fig.~\ref{fig:fig05}C, (yet it could be excited for perpendicular uniform fields). For the Oersted field, however, the actual field orientations acting on the two ferromagnetic layers are opposite, see Fig.~\ref{fig:fig05}B, and, therefore, a direct excitation of the acoustic mode is possible from symmetry arguments. The above considerations are valid for the quasi-uniform precession case, but not necessarily for the excitation of short-wavelength spin waves. In the uniform in-plane field case (and the enhanced STT case), wave excitation requires the presence and gyration dynamics of the vortex cores, that can be driven by in-plane fields and then act as kind of a perpendicular nano-stirrer, leading to a spiraling wave pattern\cite{wintz2016magnetic, sluka2019emission}. For the Oersted field case, the situation is different. Here, the vortex cores are not strictly required for the excitation of waves. In fact, simulations with artificially removed cores show comparable results to those conducted with the cores present, as shown in Supplementary Fig.~\ref{fig:figs09}. Consequentially, there is no spiraling signature in the resulting emission pattern, typically originating from vortex core gyration. The reason why nevertheless a finite wavelength is excited, lies in the very strong dispersion nonreciprocity of the acoustic Damon-Eshbach mode in the SFi \cite{wintz2016magnetic,Albisetti2020}. Here, nonreciprocity refers to the fact that counterpropagating waves of the same frequency have significantly different wavelengths (by a factor of the order of 10). Therefore, the global Oersted field can excite a quasi-homogeneous precession of the order of half the structure size (radially), and upon lateral radial reflection, this dynamics appears as a short-wavelength branch propagating away from the cores (or the rim for the other antiparallel relative vortex circulation combination possible). We refer to this effect as nonreciprocal excitation. Note that in the simulations there is also a dependence of the spin-wave amplitude on the azimuthal angle, exhibiting a two-fold symmetry. In particular, there appears a quasi-nodal line, crossing the disk center from left to right. This effect stems from the azimuth-dependent resulting torque of the form $\propto (-\mathbf{M} \times \mathbf{H}^{\mathrm{Oersted}})$, when considering the Oersted field geometry and the vortex magnetization.  Figure~\ref{fig:fig05}B illustrates the torque generated by the Oersted field, and Fig.~\ref{fig:fig05}C depicts the torque resulting from the uniform external field from a stripline antenna. The possible extent to which the spin-texture dynamics (vortex core gyration, domain wall oscillation) still contributes to the excitation of waves in parallel to the nonreciprocal oersted-field excitation effect remains a challenge to be quantified.

\textbf{Direction-steerable spin-wave emission.} In the above experiments and simulations, the spin-wave emission pattern was fixed by the specific geometries of the spin texture and the current flow. In the following, we will demonstrate a proof-of-principle system where the emission pattern and, in particular, its directional properties can be tuned by the drive current. To this end, we used another SFi disk of the same diameter (9 $\mu$m) made of a $\mathrm{Ni_{81}Fe_{19}}$(\SI{44.9}{nm})/Ru(\SI{0.8}{nm})/$\mathrm{Co_{40}Fe_{40}B_{20}}$(\SI{46.6}{nm}) stack (Sample \#2), which is illustrated in Supplemental Fig.~\ref{fig:figs04}. $\mathrm{Co_{40}Fe_{40}B_{20}}$ is well-known to be a magnetostrictive material \cite{Joule1847,Lee1955}, i.e., it changes its anisotropy with mechanical strain through the magneto-elastic effect. 

\begin{figure*}
	\centering
	\includegraphics[width=\textwidth]{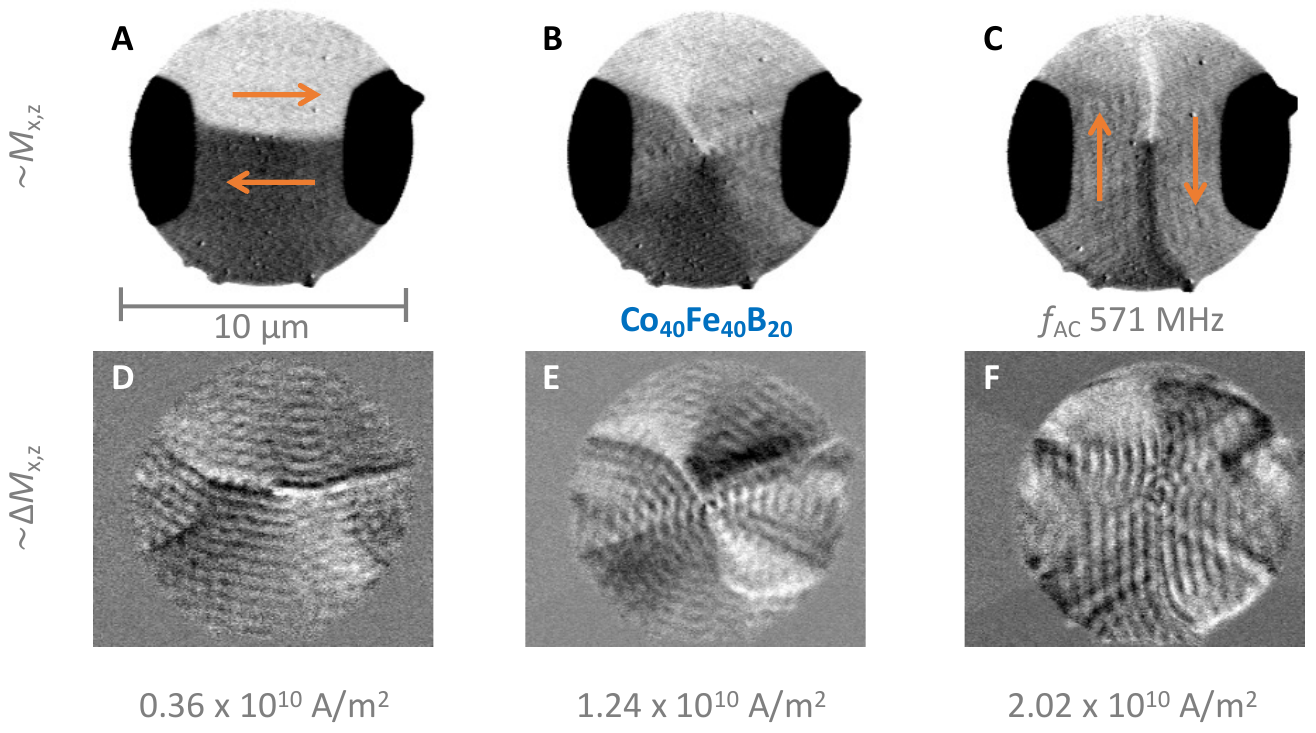}
	\caption{\textbf{TR-STXM imaging of direction-steerable spin-wave emission in SFi Sample \#2.} Response to alternating currents of $f_{AC} = \SI{571}{MHz}$ with current densities as provided below the individual columns. Images recorded at the Co $\mathrm{L_{3}}$ edge. Top row (A,B,C), direct absorption snapshots showing both topographic and magnetic contrast, the latter with mixed sensitivity ($\sim M_{x,z}$). Bottom row (D,E,F), normalized snapshots that highlight the magnetic dynamics ($\sim \Delta M_{x,z}$). The orange arrows indicate the magnetic background orientation.}
 \label{fig:fig06}
\end{figure*}
 
 %The ac current heats up the copper contacts that elongate the SAF stack leading to magnetostriction which elongates the magnetic vortex as shown in A if the amplitude of the current sent through the SAF is $I_{\mathrm{eff}} = \SI{3.6}{mA}$. The domain wall shown in B. rotates nearly  $\SI{45}{\deg}$ if the amplitude is increased to $I_{\mathrm{eff}} = \SI{7.3}{mA}$, whereas the domain wall rotates $\SI{90}{\deg}$ if the excitation amplitude is $I_{\mathrm{eff}} = \SI{11.7}{mA}$. In D-F one can see that the spin-wave propagation direction can be tuned with the applied current due to the rotation of the domain wall, and magnetic vortex as the spin-wave emitter.} 

Figure~\ref{fig:fig06} A shows a STXM image with partial in-plane sensitivity of the $\mathrm{Co_{40}Fe_{40}B_{20}}$ layer of Sample \#2. Although excited by an alternating current, the overall magnetic state is analogous to the equilibrium state of the sample. As a consequence of an intrinsic uniaxial magnetic anisotropy along the $x$-axis (horizontal), the vortex circulation is broken up into two oppositely oriented domains as indicated by the orange arrows. These domains are connected by 180${^\circ}$ domain walls that have a vortex core remaining in the center \cite{sluka2019emission}. The origin of the intrinsic uniaxial anisotropy presumably stems either directly from the film growth or from a potential tensile strain exerted by the copper leads. As for Sample \#1, the Ru mediated interlayer exchange coupling enforces an antiparallel alignment of the magnetic in-plane components between the two ferromagnetic layers of Sample \#2, as it can be seen in Supplemental Fig.~\ref{fig:figs04}. 

In fact, Fig.~\ref{fig:fig06}A shows the response of the sample as TR-STXM snapshots with mixed in-plane and out-of-plane sensitivity ($\sim M_{x,z}$) towards an alternating current of $f_{AC} = \SI{571}{MHz}$ flowing through the sample at $j = \SI{0.36e10}{\mathrm{Am^{-2}}}$ amplitude. It shows the direct absorption view, where magnetic and nonmagnetic features are visible (the disk itself, the copper leads, the magnetic domains, and some signature of spin waves). Figure~\ref{fig:fig06}D shows the corresponding snapshot as normalized view, with sensitivity to magnetic changes over time compared to the time-average state ($\sim \Delta M_{x,z}$).
In this view, we see a clear emission pattern of rather plane spin waves from the horizontal domain walls into both directions (up and down). There are also oscillations of the domain wall itself visible with some additional complexity in the regions outside the center area, below the leads (possible wall branching). For the animated version of Fig.~\ref{fig:fig06}, we refer to Supplemental Movie M5. When the current density is increased to $j = \SI{1.24e10}{Am^{-2}}$, we see in Figs.~\ref{fig:fig06}B,E that the time average state has changed compared to the low-current/static state. The earlier well-defined horizontal domain walls have changed into a more continuous vortex state, with some diagonal wall features remaining. As one can see in the normalized snapshot in Fig.~\ref{fig:fig06}E, the spin-wave emission pattern is now more isotropic than before. When the current density increases further to $j = \SI{2.02e10}{Am^{-2}}$, the time averaged state changes again, as shown in Figs~\ref{fig:fig06}C,F. There is a reappearance of well defined $180^{\circ}$ domain walls, however, this time they are oriented along the $y$-axis (vertical). The resulting spin-wave emission pattern in Fig.~\ref{fig:fig06}F mainly consists of plane waves again emitted from the domain walls, now propagating along the horizontal axis. By tuning the excitation current density, we can control the orientation of the domain walls in Sample \#2 to be horizontal or vertical, which in turn gives us control of the axis along which plane spin waves are emitted. For intermediate current densities, the system assumes a less anisotropic state, with almost isotropic spin-wave emission. At the same time, the spin-wave wavelength appears to be almost unaffected by this way of steering the emission direction.
\begin{figure*}
	\centering
	\includegraphics[width=0.9\textwidth]{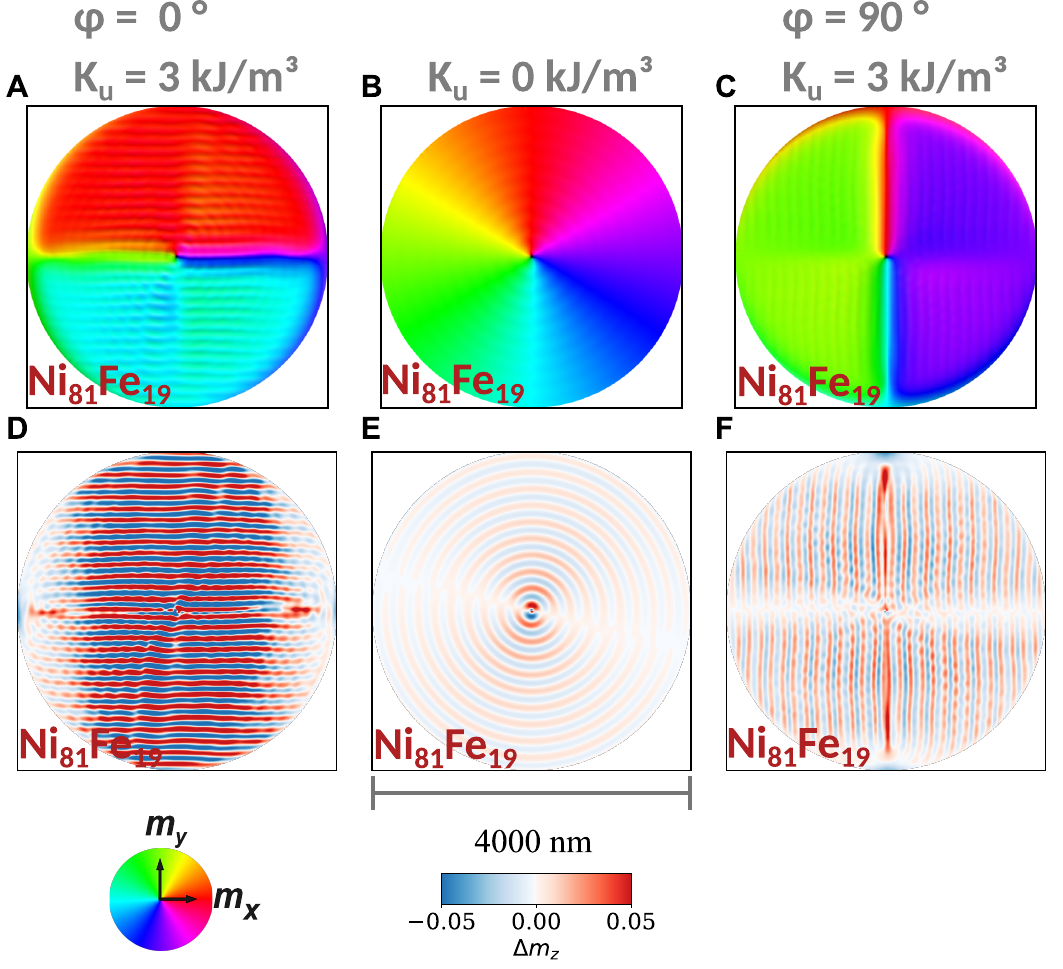}
	\caption{\textbf{Direction-steerable spin-wave emission in anisotropic SFi vortex pairs (simulation)} (A) magnetization of the bottom $\mathrm{Ni_{81}Fe_{19}}$ layer as color-coded snapshot, where the horizontal in-plane domains with the vortex core are visible, as well as the modulation in the magnetization due to the spin-wave amplitude, which is depicted more clearly in (D) as $\Delta m_z$ snapshots. In A and D, the strain-induced uniaxial anisotropy in the $\mathrm{{Co_{40}Fe_{40}B_{20}}}$ layer was set to be parallel to the applied current ($\varphi = 0$) with an anisotropy constant of $K_u = \SI{3}{kJ/m^3}$. This anisotropy was set to zero in B and E, to simulate the current/strain-induced compensation of the anisotropy. To simulate the rotation of the anisotropy, leading to the emission of spin waves in the vertical direction, the anisotropy axis was rotated to align along the $y$ axis ($\varphi = 90 ^\circ$) in C and F.}
 \label{fig:fig07}
\end{figure*}

The most likely explanation for the reorientation of the domain walls lies in a combination of joule heating and magneto-elastic effects (inverse magnetostriction). With increasing current density, the disk starts to warm up; urging it to expand. Here, the copper leads provide a certain constraint for a horizontal expansion, which exerts essentially a compressive strain on the disk. In the course of this action, the initially horizontal anisotropy is first reduced, and eventually realigned to a vertical orientation. This explanation is in line with the positive magnetostrictive constant of $\mathrm{Co_{40}Fe_{40}B_{20}}$ \cite{Wang2005,Finizio2018}, while the magnetosrictive effects in $\mathrm{Ni_{81}Fe_{19}}$ are negligible \cite{McKeehan1926}. It is also consistent with the fact that the anisotropy control is rather independent of the driving frequency, as can be seen in the Supplemental Fig.{~\ref{fig:figs05} and Supplemental Movie M6, showing similar state responses at $f_{AC} = \SI{71}{MHz}$ and $f_{AC} = \SI{1071}{MHz}$. 
We underpin this hypothesis by further micromagnetic simulations, where we model Sample \#2 with dimensions equal to those used for Sample \#1. To this end, at first we  introduce an in-plane anisotropy to the effective field along the $x$-axis. We demonstrate in Supplemental Fig.~\ref{fig:figs08} that by increasing the anisotropy constant $K_u$, a bidomain state is induced, similar to that observed in our experimental findings. Then, we simulate the current-driven Oersted field excitation to model the corresponding spin-wave generation. In Fig.~\ref{fig:fig07}, we depict three cases. First, the anisotropy axis is chosen to be along the $x$-axis and parallel to the applied current (Figs.~\ref{fig:fig07}A and D), where $K_u = \SI{3}{kJ/m^3}$. Figure~\ref{fig:fig07}D shows a clear anisotropic emission of plane spin waves, originating at the domain wall. By that we reproduce the low current density regime in Fig.~\ref{fig:fig06} D. If we assume that upon the intermediate increase of the current density, the resulting magneto-elastic effect compensates for the original anisotropy, then a regular vortex pair is obtained again as shown in Fig.~\ref{fig:fig07}B, where the spin-wave emission is now rather isotropic, as can be seen in Fig.~\ref{fig:fig07}E. Finally, when the magneto-elastic effect induces the domain wall to rotate by $90^\circ$, the emission of spin waves occurs collinear to the current direction (Fig.~\ref{fig:fig07}C,F). We provide the animations analogous to Fig.~\ref{fig:fig07} in the Supplemental Movie M7.

Note that while there are certainly clear benefits in being able to control the spin-wave emission direction by the drive current, there are also other, potentially more practical ways to control strain/anisotropy directly, e.g. by piezoelectrics or voltage controlled magnetic aniostropy (VCMA). In addition to direction-steerable emission, Sample \#2 can also be used to demonstrate single-pulse driven excitation of individual spin-wave wave packages (see Supplemental Fig.~\ref{fig:figs07} and Supplemental Movie M8). This effect further highlights the agility of current-induced spin-wave excitation in SFi, neither requiring particular resonance frequencies nor multiple onset excitation cycles.

\section*{Discussion}
We demonstrated efficient current-driven excitation of spin waves in an antiferromagnetically coupled magnetic vortex pair. By employing TR-STXM, we showed that the spin waves are generated when the SFi is subjected to an alternating current flowing directly through it. Micromagnetic simulations revealed that the Oersted field generated in the SFi by the current is the main contributor to the emission of spin waves in the magnetic vortex pair. By further comparing this to the contributions from both STT and the uniform magnetic field of a stripline antenna, we determined that the excitation efficiency of the current-driven Oersted field in terms of energy is at least three orders of magnitude higher than the other two methods investigated. To demonstrate an additional advancement of current-driven spin-wave generation, we prepared the SFi from a magnetostrictive material, where initial strain induces an additional anisotropy. This additional anisotropy distortes the vortex state, leading to the formation of a bidomain state with a pronounced domain wall in between the domains in both magnetic layers. We demonstrated that the propagation direction of plane spin waves excited in this system can be redirected if the amplitude of the applied current is varied, allowing even for a $90^{\circ}$ change in the emission direction. Our findings represent a significant advance in the current-driven, energy-efficient excitation and control of spin waves for potential magnonic nanodevices.

\section*{Materials and Methods}

\textbf{Sample Preparation}

Analogous to Ref.
\cite{sluka2019emission} and \cite{wintz2016magnetic}, the magnetic multilayers were deposited by means of magnetron sputtering onto X-ray transparent silicon-nitride window membranes. For oxidation protection, an aluminium capping layer of 3 nm thickness was used. The microdisks were fabricated by means of electron beam lithography (EBL) and consecutive ion beam etching. After an initial oxygen plasma treatment for adhesive purposes, a negative resist (MA-N 2910) was spun onto the films and baked out. In a second step, the microdisks were exposed by EBL and the samples were then developed for 300 s in MA-D 525 and subsequently rinsed in deionized water (60 s). Finally, the samples were milled using an argon ion beam at two different angles (85° and 5°) with endpoint detection for approximately one hour to physically etch the magnetic microdisks out of the continuous films. An acetone bath (12 hours) and a second oxygen plasma step (20 minutes) were applied to remove the remaining resist. For being able to apply electric currents flowing laterally through the samples, copper/aluminium leads of 200/5 nm thickness were fabricated to overlap at about 2 $\mu$m with a microdisk at two opposing rim positions (see Fig. \ref{fig:fig01}). For this purpose EBL, electron beam evaporation and lift-off processing was used. The resulting resistances of the samples were typically between 10 $\Omega$ and 100 $\Omega$.

\textbf{Scanning-Transmission X-ray Microscopy}

To image the magnetic orientation in the multilayer disks, synchrotron-based scanning-transmission X-ray microscopy (STXM) was employed \cite{weigand2022timemaxyne}, as schematically depicted in Fig.~\ref{fig:fig01}A. For STXM, a Fresnel zone plate is used to focus monochromatic X-rays onto the sample, while undiffracted X-rays and those of higher diffraction order are blocked by a combination of the zone plate center stop and a circular order selecting aperture. The X-ray intensity transmitted through the sample is collected by a point detector. To get an image, the sample is raster scanned through the focused beam, yielding a lateral resolution of approximately 25 nm. Magnetic contrast is achieved by exploiting the X-ray magnetic circular dichroism effect (XMCD) \cite{schutz1987absorption}, occurring at the element specific resonant absorption edges. For the given magnetic multilayer of Sample \#1, therefore, the two layers can be separately imaged in terms of magnetic orientation by measuring at photon energies of Co L3 $\sim$778 eV, and Fe L3 $\sim$708 eV or Ni L3 $\sim$853 eV, respectively. For Sample \#2, on the other hand, Co and Ni photon energies lead to layer-selective images, while the Fe energy provides an integrated signal of both layers. In general, the logarithmic transmission contrast detected is proportional to the projection of the magnetization on the X-ray propagation direction, which means that in normal incidence XMCD is sensitive to the perpendicular magnetization component, while inclination of the sample provides additional information on in-plane magnetization components.

The spin-wave dynamics in the multilayer disks was imaged using pump-and-probe time-resolved STXM (TR-STXM) \cite{weigand2022timemaxyne}. For TR-STXM, the inherent time structure of the incident X-ray pulses (500 MHz repetition rate and $\sim$100 ps effective pulse width) is used. The accessible sinusoidal excitation frequencies are of the form $f_{M}=M/N*500 \mathrm{MHz}$ where $M$ is an integer multiplier and $N$ corresponds to the integer number of phases acquired simultaneously. The level of the excitation signal was measured both before and after the sample (via a pick-off tee) by means of an oscilloscope. The mean of the input and transmitted signal was used to estimate the resulting current density in the sample by assuming the diameter and height of the disk as relevant areas for the current cross section.

\textbf{Micromagnetic Simulations}.

We use \texttt{magnum.np}\cite{bruckner2023magnumnp}, an open-source GPU-accelerated micromagnetic simulation library based on PyTorch, to conduct our numerical investigations. We modeled the SFi stack in a total simulation box of $1000\times 1000\times 20$ cells with discretization lengths $(\SI{4}{nm}, \SI{4}{nm}, \SI{4.5}{nm} )$, where the first ten cells along the dimension $z$ are chosen to be the bottom layer and the other ten cells represent the top layer. For Sample \#1, we use the following material parameters $M_s^{Co} = \SI{1240}{kA/m}$, $A^{Co} = \SI{16}{pJ/m}$, and $K_u^{Co} = 0$ for the bottom Co layer, as well as $M_s^{\mathrm{Ni_{81}Fe_{19}}} = \SI{750}{kA/m}$, $A^{\mathrm{Ni_{81}Fe_{19}}} = \SI{7.5}{pJ/m}$, and $K_u^{\mathrm{Ni_{81}Fe_{19}}} = 0$, for the top $\mathrm{Ni_{81}Fe_{19}}$ layer, in agreement with previous work and literature\cite{wintz2016magnetic, sluka2019emission}. Each layer is modeled as a disk with a diameter of $\SI{4}{\mu m}$, where cells outside the disk are initialized with $M_s, A, K_u = 0$, as usual in finite-difference-based micromagnetic simulations. We use second-order Ruderman-Kittel-Kasuya-Yosida (RKKY) coupling to ensure the correct antiferromagnetic coupling between cells\cite{suess2023accurate}. The coupling strength is chosen to be $J_{\mathrm{RKKY}} = \SI{-0.1}{mJ/m^2}$. For Sample \#2, the material parameters were chosen as $M_s^{\mathrm{Co_{40}Fe_{40}B_{20}}} = \SI{1240}{kA/m}$, $A^{\mathrm{Co_{40}Fe_{40}B_{20}}} = \SI{12}{pJ/m}$, and $K_u^{\mathrm{Co_{40}Fe_{40}B_{20}}}$ was varied for the top  $\mathrm{Co_{40}Fe_{40}B_{20}}$ layer. We used $M_s^{\mathrm{Ni_{81}Fe_{19}}} = \SI{750}{kA/m}$, $A^{\mathrm{Ni_{81}Fe_{19}}} = \SI{8}{pJ/m}$, and $K_u^{\mathrm{Ni_{81}Fe_{19}}} = 0$, for the bottom $\mathrm{Ni_{81}Fe_{19}}$ layer. The Gilbert damping is $\alpha = 0.01$. In all cases, we assume an artificially increased damping region of $\alpha = 0.1$ for the annulus of the outermost 50 nm. This allows us to avoid reflections of the spin waves at the rim of the sample and thus to achieve the steady state much faster. 

We begin our simulations by parametrizing a SFi with two magnetic vortices of opposite circulations but parallel polarities of the vortex cores. We relax the magnetization state by numerically integrating the Landau-Lifshitz-Gilbert (LLG) equation at a high Gilbert damping parameter of $\alpha = 1$ for $\SI{20}{ns}$, where

\begin{equation}
	\centering
	{\dfrac{\partial\mathbf{m}}{{\partial}t}} = {-\gamma\mathbf{m}\times{\mathbf{H}}^{\mathrm{eff}}+\alpha{\mathbf{m}}\times{\dfrac{\partial\mathbf{m}}{{\partial}t}}},
	\label{eq:LLG}
	\end{equation}
describes the temporal evolution of the magnetization vector field. Here, $\gamma = \SI{2.2128e5}{m/As}$ is the gyromagnetic ratio, and $\mathbf{H}^{\mathrm{eff}}$ is the effective field that can be derived as the variational derivative of the different relevant energetic contributions. 
For the statical relaxation process of the total energy, the effective field includes only contributions from the demagnetization, exchange, uniaxial anisotropy, and interlayer exchange (RKKY) energies. Note that we do not use a physical spacing layer, but a second-order finite-difference implementation of the RKKY~\cite{suess2023accurate}. This will slightly change the stray fields in between the two magnetic layers in comparison to the experiments, but it suffices to reproduce the most relevant effects. Once the magnetization is relaxed, we include additional dynamic contributions such as the current-induced Oersted field, the alternating uniform Zeeman field, or the spin-transfer torque (STT). Note that all terms are already implemented in \texttt{magnum.np} and can be used together or individually as desired.

\textbf{Oersted Field.} The charge current that passes through a conducting material, such as we consider in our experiments and modeling, will generate an Oersted field that can be calculated by means of the Biot-Savart law, where 

\begin{equation}
  \mathbf{H}^{\mathrm{Oersted}}(\mathbf{x}, t) = \frac{1}{4\pi} \int \mathbf{j}(\mathbf{x'}, t) \times \frac{(\mathbf{x} - \mathbf{x'})}{\lvert \mathbf{x} - \mathbf{x'} \rvert^3} \, \mathrm{d}\mathbf{x'},
\end{equation}
where $\mathbf{j}(\mathbf{x'}, t)$ is the space and time dependent current density flowing through the conductor. In our modeling we assume that the current density is homogenous, and thus, a spatially uniform and constant vector that is only time-dependent. This equation is solved in the Fourier space, similar to the calculation of the demagnetization field, using a Fast Fourier-Transform algorithm, which provides a substantial speed-up. For simulations where we consider the current-driven excitation of spin waves, the Oersted field is solved for each time step according to the adaptive time-stepped Runge-Kutta-Fehlberg time integrator provided as default in \texttt{magnum.np.} For the stripline antenna simulations, we precompute the Oersted field at a constant current density $j = \SI{1e9}{Am^{-2}}$, and take the maximum of the $y$ component of the calculated Oersted field as the value for the spatially homogeneous, stripline generated field.

\textbf{Spin-Transfer Torque.}
The spin of the elecrons flowing through our magnetic stack can be polarized due to the magnetization. Such spin-polarized electrons will exert a torque on the magnetization, called the spin-transfer torque. This additional, current-induced torque can be modeled in a continuum theory by using the Zhang and Li approach\cite{abert2019micromagnetics}, where the STT is modeled as an additional torque in the Landau-Lifshitz-Gilbert equation mentioned above, and hence becoming the Landau-Lifshitz-Gilbert-Slonczewski equation, where
\begin{equation}
	\centering
	{\dfrac{\partial\mathbf{m}}{{\partial}t}} = {-\gamma\mathbf{m}\times{\mathbf{H}}^{\mathrm{eff}}+\alpha{\mathbf{m}}\times{\dfrac{\partial\mathbf{m}}{{\partial}t}}+\mathbf{T}_{\mathrm{STT}}}.
	\label{eq:LLS}
	\end{equation}
 
 The additional torque
\begin{equation}
    T_{\mathrm{STT}} = -b \times \mathbf{m} \times \left[ \mathbf{m} \times \left( \mathbf{j}_e \cdot
\boldsymbol{\nabla} \right) \mathbf{m} \right] -b \xi \mathbf{m} \times \left( \mathbf{j}_e \cdot
\boldsymbol{\nabla} \right) \mathbf{m},
\label{eq:stt}
\end{equation}
describes how the STT acts due to the local gradient of the magnetization, where $\xi$ is the degree of nonadiabacity, and $b$ is described as

\begin{equation}
    b = \dfrac{\beta \mu_B}{eM_s(1 + \xi^2)}.
\end{equation}

The polarization rate of the conducting electrons, which possess an elementary charge of $e$, is symbolized by $\beta$. The Bohr magneton is denoted by $\mu_B$. For all simulations involving STT, we use $\xi = 0.05$ and $b = \SI{72.17e-12}{}$.

% Produces the bibliography via BibTeX.
\bibliography{manuscript}

\textbf{Acknowledgements:} The computational results presented have been achieved, in
part, using the Vienna Scientific Cluster (VSC). F.B. and D.S. gratefully acknowledge the Austrian Science Fund (FWF) for support through Grant No. I 4917 (MagFunc). S.K. and C.A.
gratefully acknowledge the Austrian Science Fund (FWF) for support through Grant No. P34671 (Vladimir). S.K. and D.S. acknowledge the Austrian Science Fund (FWF) for support
through Grant No. I 6267 (CHIRALSPIN). We thank Helmholtz-Zentrum Berlin for the allocation of synchrotron radiation beamtime. We thank Roland Mattheis (IPHT Jena) for the sputter deposition of the magnetic multilayers, Sina Mayr for fruitful discussions. Support by the Nanofabrication Facilities Rossendorf at IBC is gratefully acknowledged.

\textbf{Author Contributions:} S.K., C.A., D.S., and S.W. conceived the project. S.K., F.B., C.A., and D.S wrote and improved the micromagnetic code. S.K. performed the micromagnetic simulations. S.K. and S.W. analyzed the simulation data. K.S. and S.W. patterned the samples. K.S., M.W., and S.W. performed the measurements. K.S., and S.W. analyzed the experimental data. S.K. and S.W. wrote the initial manuscript with input from all authors. All authors have reviewed and commented on the final manuscript.

\textbf{Competing Interests:} The authors declare that they have no competing interests.

\textbf{Date and materials availability:} All data needed to evaluate the conclusions in the paper are present in the paper and/or the Supplementary Materials. Additional data can be requested from the corresponding author upon request.

\end{document}